\documentclass[preprint,times]{aastex631}

\usepackage[T1]{fontenc}
\usepackage{savesym}
\savesymbol{tablenum}
\usepackage{graphicx}
\usepackage{xcolor}
\usepackage{siunitx}
\usepackage{amsmath}
\usepackage{tabularx}
\usepackage{graphicx}
\usepackage[figuresright]{rotating}
\usepackage{color}  
\usepackage{subfiles} 

\newcommand{\K}{\unit{K}}
\newcommand{\kms}{\ensuremath{{\rm\, km\,s^{-1}}}}

\usepackage{CJKutf8}
\received{\today}

\begin{document}
\begin{CJK*}{UTF8}{gbsn}
\title{A closer look at the origin of LINER emission and its connection to evolved stars with a machine learning classification scheme}

\author[0000-0002-9220-0039]{Ahmad Nemer}
\affiliation{New York University Abu Dhabi, PO Box 129188, Abu Dhabi, UAE}
\affiliation{Center for Astrophysics and Space Science (CASS), New York University Abu Dhabi, PO Box 129188, Abu Dhabi, UAE}

\author[0000-0002-6425-6879]{Ivan Yu. Katkov}
\affiliation{New York University Abu Dhabi, PO Box 129188, Abu Dhabi, UAE}
\affiliation{Center for Astrophysics and Space Science (CASS), New York University Abu Dhabi, PO Box 129188, Abu Dhabi, UAE}
\affiliation{Sternberg Astronomical Institute, Lomonosov Moscow State University, Universitetskij pr., 13,  Moscow, 119234, Russia}

\author[0000-0003-4679-1058]{Joseph D. Gelfand}
\affiliation{New York University Abu Dhabi, PO Box 129188, Abu Dhabi, UAE}
\affiliation{Center for Astrophysics and Space Science (CASS), New York University Abu Dhabi, PO Box 129188, Abu Dhabi, UAE}
\affiliation{Center for Cosmology and Particle Physics, New York University, 726 Broadway, room 958, New York, NY 10003}

\author[0000-0002-9879-1749]{Changhyun Cho}
\affiliation{New York University Abu Dhabi, PO Box 129188, Abu Dhabi, UAE}
\affiliation{New York University, 726 Broadway, New York, NY 10003}




\begin{abstract}
Identifying the dominant ionizing sources in galaxies is essential for understanding their formation and evolution. Traditionally, spectra are classified based on their dominant ionizing source using strong emission lines and Baldwin, Phillips, \& Terlevich (BPT) diagrams. The ionizing source is traditionally determined by the emission line ratios using the BPT diagrams. Low-Ionization Nuclear Emission-line Regions (LINERs) are a class of ionizing mechanisms that is observationally identified but with a poorly understood origin, unlike the case of star forming regions and active galactic nuclei (AGN).  LINERs, typically found in early-type galaxies, are often associated with low-luminosity AGN activity but may also be powered by aging stellar populations, particularly post-Asymptotic Giant Branch (p-AGB) stars.  In this study, we employ a machine-learning-based encoder, Spender, to analyze the full MaNGA IFU spectra and identify key spectral features of LINERs. By examining the continuum and line emission of these spaxels, our approach aims to uncover hidden patterns and better understand the dominant ionizing sources. We show in this work that the neural network-based encoder was able identify LINER sources from the stellar continuum alone. The characteristics of the stellar population underlying LINER regions are consistent with evolved low mass stars implying that the source driving LINER emission is probably p-AGB stars rather than AGN activity. 
\end{abstract}



\section{Introduction}\label{sec:intro}
Determining the energy production and transmission mechanisms within a galaxy is required to understanding their formation and evolution. Powerful, luminous sources produce ionizing photons, and therefore play a key role in shaping their host galaxies. Baldwin, Phillips and Terlevich \citep[hereafter BPT]{BPT+1981} showed that it is possible to differentiate between gas ionized by emission from type 2 active galactic nuclei (AGN) as opposed to Star Forming (SF) regions by analyzing the intensity ratios of two sets of relatively strong emission lines; this method was further improved by \citet{Veilleux+1987}(e.g. Fig.~\ref{1}). The bottom left part of these diagrams consist of SF galaxies, while the top right part is attributed to galaxies with an AGN \citep{Kauffmann+2003} requiring a hard spectrum of ionizing photons to excite. The AGN part has been subdivided into an upper branch called Seyfert (Sy), and a lower branch termed LINER \citep{Kewley+2006}; LINER stands for ‘Low-Ionization Nuclear Emission Regions’. 

Unlike galaxies dominated by SF regions or Seyferts, which have line ratio patterns that clearly indicate their ionizing sources as either young massive stars or AGN, LINER emission can be explained by a variety of ionization processes. These include photoionization by low luminosity AGN \citep{Ferland+1983,Halpern+1983}, post-AGB stars \citep{Binette+1994}, fast radiative shocks \citep{Dopita+1995}, accretion of warm gas in cooling flows \citep{Heckman+1981}, or a mixture of the above. Consequently, the precise ionization mechanism for LINERs is currently a topic of considerable debate. Some LINERs clearly exhibit evidence for containing low-luminosity AGNs, as identified by a central non-stellar source (e.g. an X-ray or UV point source, or a compact radio core) or signs of AGN accretion (double-peaked broad H emission) \citep{Filippenko+2003}, but many others do not. 

In an early study, \citet{Binette+1994} showed that post-AGB stars emit enough hard ionizing radiation to explain the small but detectable H$\alpha$ luminosities and equivalent widths (EWs) observed in some early-type galaxies. \citet{Stasinska+2008} introduced the idea of ``retired'' galaxies, which have little ongoing star formation, and used stellar population synthesis modeling to demonstrate that these galaxies can produce detectable optical line emission due to post-AGB stars and hot white dwarfs. This could account for a significant portion of LINER-type galaxies without requiring AGN-driven excitation. \citet{Fernandes+2011} expanded on this concept and applied it to a large sample of galaxies with central region spectroscopy available from the Sloan Digital Sky Survey (SDSS). They used a diagnostic diagram involving $H\alpha$ EW and [N\,\textsc{ii}]/$H\alpha$ line ratios to classify sources into SF, AGN, "retired," and completely passive categories, identifying a bimodal distribution in $H\alpha$ EW that separates AGN from evolved star (assumed to be post-AGB) ionization, with a threshold at 3 \AA. More recently, \citet{Belfiore+2016} analyzed data from 646 galaxies in the SDSS-IV MaNGA Integral Field Unit (IFU) survey and \citet{Singh+2013} did the same for 48 galaxies in the CALIFA IFU survey, both revealing extended emission several kiloparsecs from the galaxy center with LINER-type ratios in many cases. Based on these results, it was suggested post-AGB stars are the dominant ionizing source and concluded that shock-excited LINER regions are uncommon.

Early-type galaxies (ETGs) are generally thought to be gas-poor and passively evolving systems belonging to red sequence galaxies. Nevertheless, it is known that ETGs can also contain reservoirs of diffuse ionized gas \citep{Papaderos+2013,Singh+2013}. In fact, \citet{Yan+2006} found that up to 30\% of red sequence galaxies in the SDSS exhibit emission lines with ratios typical of LINERs. The low-ionization emission was originally discovered in the nuclear regions of galaxies with a lower luminosity than typical Sy galaxies while requiring a hard ionizing source to excite \citep{Filippenko+2003}. \citet{Kauffmann+2003} found that both Sy and LINER emission galaxies resemble ETGs morphologies, with the former harboring a stellar population typical of late-type (star forming) galaxies and a high luminosity AGN activity, while the latter is characterized by an old stellar population and a low luminosity AGN. These findings convinced many authors that both LINER and Sy type emission is a result of AGN activity \citep{Kewley+2006,Schawinski+2007,Kauffmann+2009}. 

Initially, the term "LINER" specifically referred to a type of galaxy nucleus. These were first detected in the nuclear spectra of nearby galaxies \citep{Heckman+1981}. Nonetheless, LINER emission can be further broken down into nuclear LINER (within a few central hundred pc) , and extended LINER-like emission (typically $\geq 1 Kpc$) with the advent of modern, IFU observations of nearby galaxies ($\leq 40 Mpc$) \citep{Sarzi+2006,Belfiore+2016, Singh+2013}. These studies suggest that a significant portion of luminous elliptical galaxies contain ionized gas, with optical emission line ratios typical of LINER. Substantial evidence was presented to argue that extended LINERs cannot be powered by low luminosity AGN activity, the strongest of which presented by \citet{Yan+2018} where they show that LINER-like galaxies are powered by an outward increasing ionization parameter (unlike a centralized source such as AGN). These arguments led some researchers to conclude that extended LINERs are not true AGNs \citep{Stasinska+2008,Fernandes+2011}, but are instead more likely driven by hot, evolved stars. In fact, \citet{Yan+2018} suggested that there is insufficient evidence whether the AGN is responsible for all of the narrow line emission within the central few hundred parsecs of nuclear LINERs, and that nuclear LINERs and extended LINERs are essentially the same population in their sample. 

It is clear that relying on strong emission lines alone is insufficient to explain the origin of LINER emission in galaxies. There has been some recent efforts to use supplemental information from IFU observations about gas kinematics and physical location to identify the ionizing source of nebular emission in galaxies \citep{Agostino+2019,Johnston+2023}. Although these works were able to distinguish SF spaxels from non-SF (shock, AGN) ones, they do not explain the origin of LINER emission and its connection to Sy emission. \textbf{We show that a significant fraction of LINER regions can be identified as such from their continuum stellar absorption features alone. Additionally, the stellar properties inferred from this continuum emission is consistent with a population that contains a large number of p-AGB stars, which corroborates the idea that LINER regions are essentially photoionized by a low-mass evolved stellar population.}

In this paper, we employ a machine learning-based auto-encoder "{\sc Spender}" \citep{Melchior+2023} to identify the dominant ionizing source in the MaNGA IFU spectra using the information contained in the full spectrum. We train {\sc Spender} with pre-classified (using their strong emission lines) spectra with the aim of learning hidden key information that was not previously captured by any of the above mentioned techniques. We perform the experiment with spectra containing both the emission lines and the stellar continuum features and another dataset that contains only the continuum stellar features to highlight the importance of the latter in the classification scheme.

\section{Data}

In this study, we utilize spectroscopic data from the Mapping Nearby Galaxies at Apache Point Observatory (MaNGA) survey \citep{Abdurro'uf2022ApJS..259...35A}, which includes IFU  \citep{Drory2015AJ....149...77D} observations obtained with the Sloan 2.5-m telescope \citep{Gunn2006AJ....131.2332G}. The MaNGA survey covers approximately 10,000 galaxies in the nearby universe, representing a broad range of stellar masses and colors \citep{Wake2017AJ....154...86W}. Each galaxy was observed with 2\arcsec\ fibers arranged in bundles of various sizes, ranging from 12\arcsec (19 fibers) to 32\arcsec (127 fibers), and covering $(1.5-2.5)\times$ the effective diameter of the target. The fiber bundles feed the BOSS spectrographs \citep{Smee2013AJ....146...32S}, which offer a wavelength range of $\lambda\lambda$3,600–10,300~\AA\ and a spectral resolution of $R\sim2000$, corresponding to an instrumental dispersion of $\sigma_\mathrm{inst.}\approx75$\kms\ at 5100\AA\ \citep{Law2016AJ....152...83L}. Each galaxy was observed with three dithered exposures to ensure full spatial coverage between the fibers \citep{Law2015AJ....150...19L, Yan2016AJ....152..197Y}.
In this project, we make use of emission line measurements from the Data Analysis Products (DAP) \citep{Westfall2019AJ....158..231W,Belfiore2019AJ....158..160B,Law2021AJ....161...52L}, extracted for individual spaxels.
We additionally employ stellar population parameters from the Pipe3D data products \citep{Sanchez2022ApJS..262...36S,Lacerda2022NewA...9701895L}.

\begin{figure}[htp]
\centering
\hspace*{-0.7in}
\includegraphics[trim={0.5in 1.9in 0.5in 1.95in},clip, width=21cm,height=7cm,angle=0]{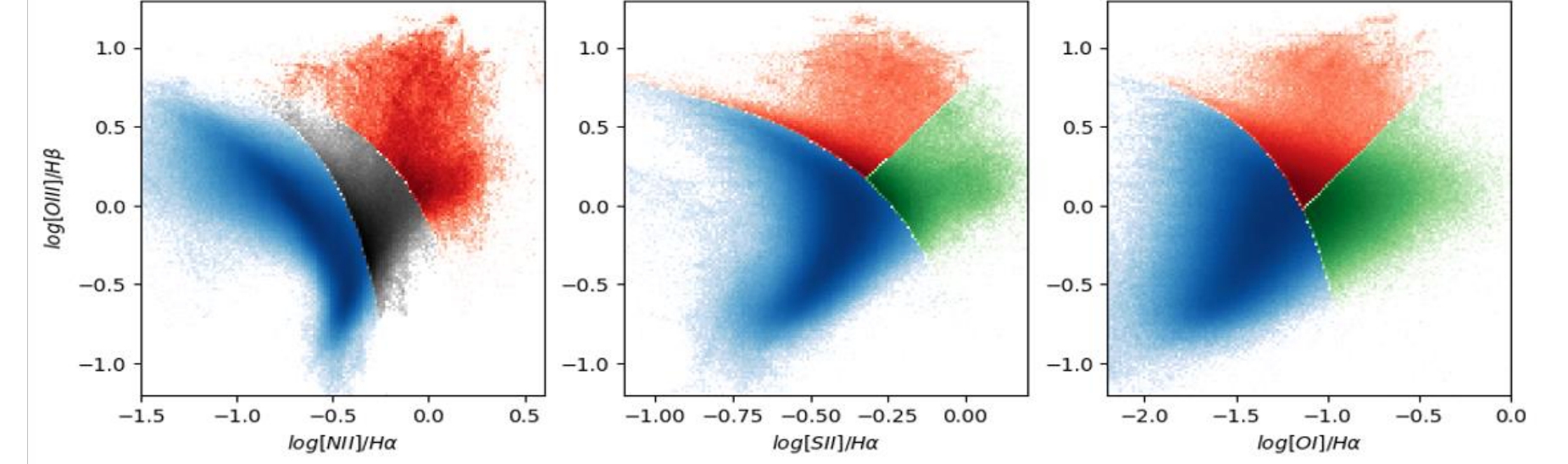}
\caption{Color coded MaNGA spaxels with S/N $\geq 5$ plotted against the BPT diagrams. The color blue represents SF spaxels, green is LINER, black is composite, and red is AGN except for the [NII] diagram where it is both AGN/LINER. }
\label{fig:fig1}
\end{figure}

\section{Methods}\label{sec:methods}

We utilize the auto-encoder {\sc Spender} \citep{Melchior+2023} to identify signals in the full MaNGA spectra that aid in galaxy classification. Auto-encoders (AEs) are feed-forward neural networks that learn to create compact representations of data without supervision. They consist of two primary components: an encoder that compresses input data into a low-dimensional latent space, and a decoder that reconstructs the original data from these latent representations. Due to their non-linear capabilities, AEs can capture complex features, such as line widths, with fewer parameters than traditional encoding methods such as principal component
analysis (PCA). Additionally, unlike line ratio diagnostics, AEs utilize continuum information in spectra, resulting in an interpretable latent space where galaxies of similar types naturally cluster \citep{Portillo+2020}.

\begin{figure}[htp]
\centering
\hspace*{-0.0in}
\includegraphics[trim={0.0in 0.0in 0.0in 0.0in},clip, width=18cm,height=10cm,angle=0]{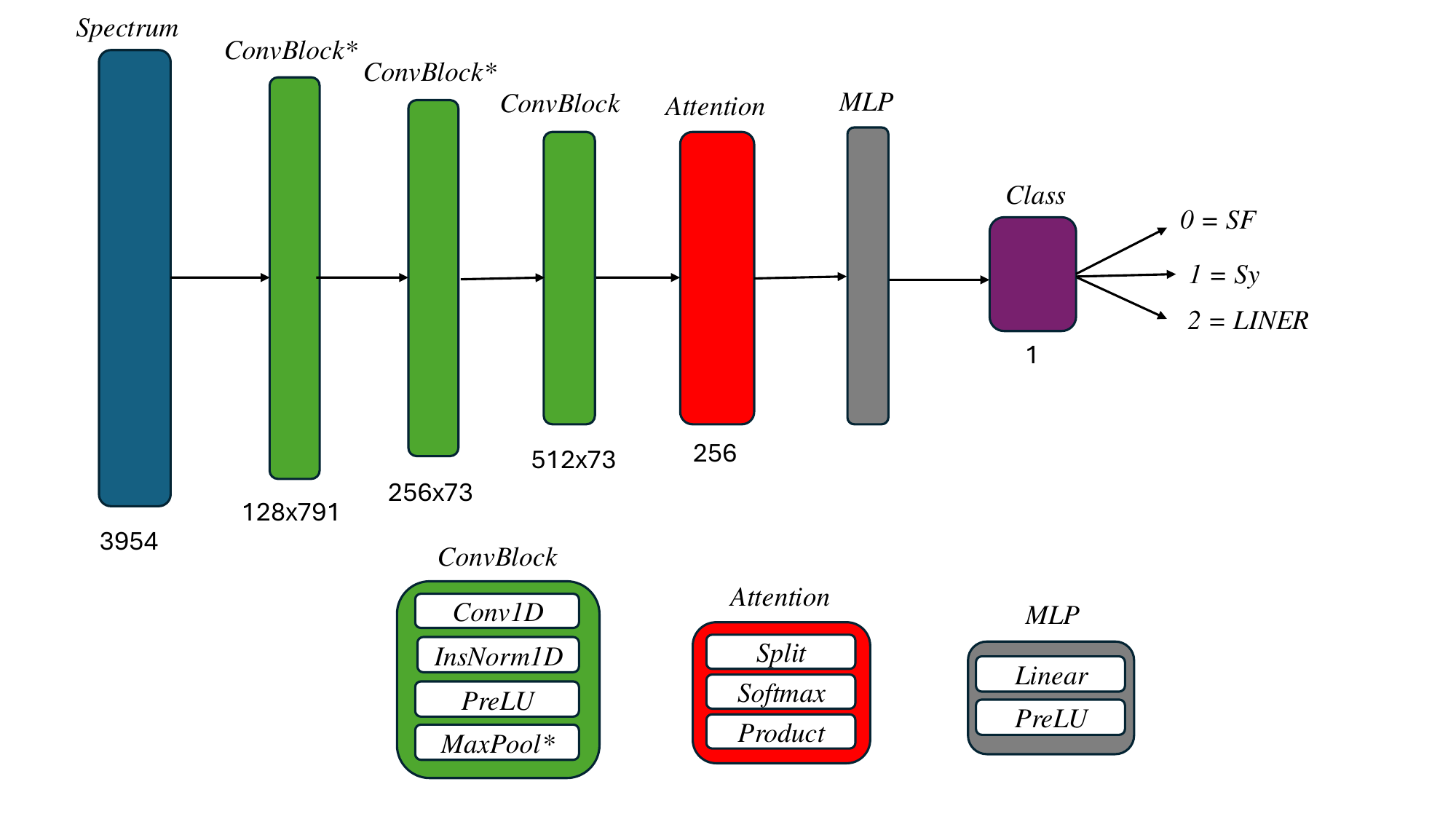}
\caption{A flow chart describing the details of the classification process using {\sc Spender}. Each type of layer has a distinct color while the input size is noted. The final layer consists of one parameter indicating the spectrum class.}
\label{fig:fig2}
\end{figure}

In this study, we employ only the encoder component of {\sc Spender}, omitting the decoder. We train it to predict the spectral class based on the dominant ionization source by minimizing the L2 norm between the input and the predicted classification parameter. This approach reduces the dimensionality of the spectra from 3,954 spectral elements to $N=1$, representing the spectral class. By training {\sc Spender} to predict the classification parameter, we design the encoder to extract features with the most significant constraining power on the dominant ionization source. Similar data compression techniques have been used in previous studies \citep{Chen2023,lemos2023,Nemer+2024}.

The encoder architecture consists of a three-layer convolutional neural network with progressively wider kernels and max-pooling layers. {\sc Spender} incorporates an attention mechanism, a widely used machine-learning technique, to identify patterns within the data \citep[for a recent review of attention methods)]{Chaudhari+2021}. This method involves learning to assign weights to input data features that are most relevant for subsequent tasks. The output vector contains these attention weights, highlighting the locations where significant signals are detected, thereby emphasizing these features before inputting them into a multi-layer perceptron to generate the latent vectors. We leverage this attention mechanism to identify which parts of the spectrum—beyond the prominent emission lines—are crucial for classification. In our work, the latent vector comprises a single parameter representing the classification parameter. We show a flow chart of the classification process using {\sc Spender} in Fig.~\ref{fig:fig2} where we report the details of each layer. For a detailed description of {\sc Spender}, we refer the reader to \citet{Melchior+2023}.

\begin{figure}[htp]
\centering
\hspace*{-0.4in}
\includegraphics[trim={3.2in 0.1in 2.9in 0.1in},clip, width=15cm,height=14cm,angle=0]{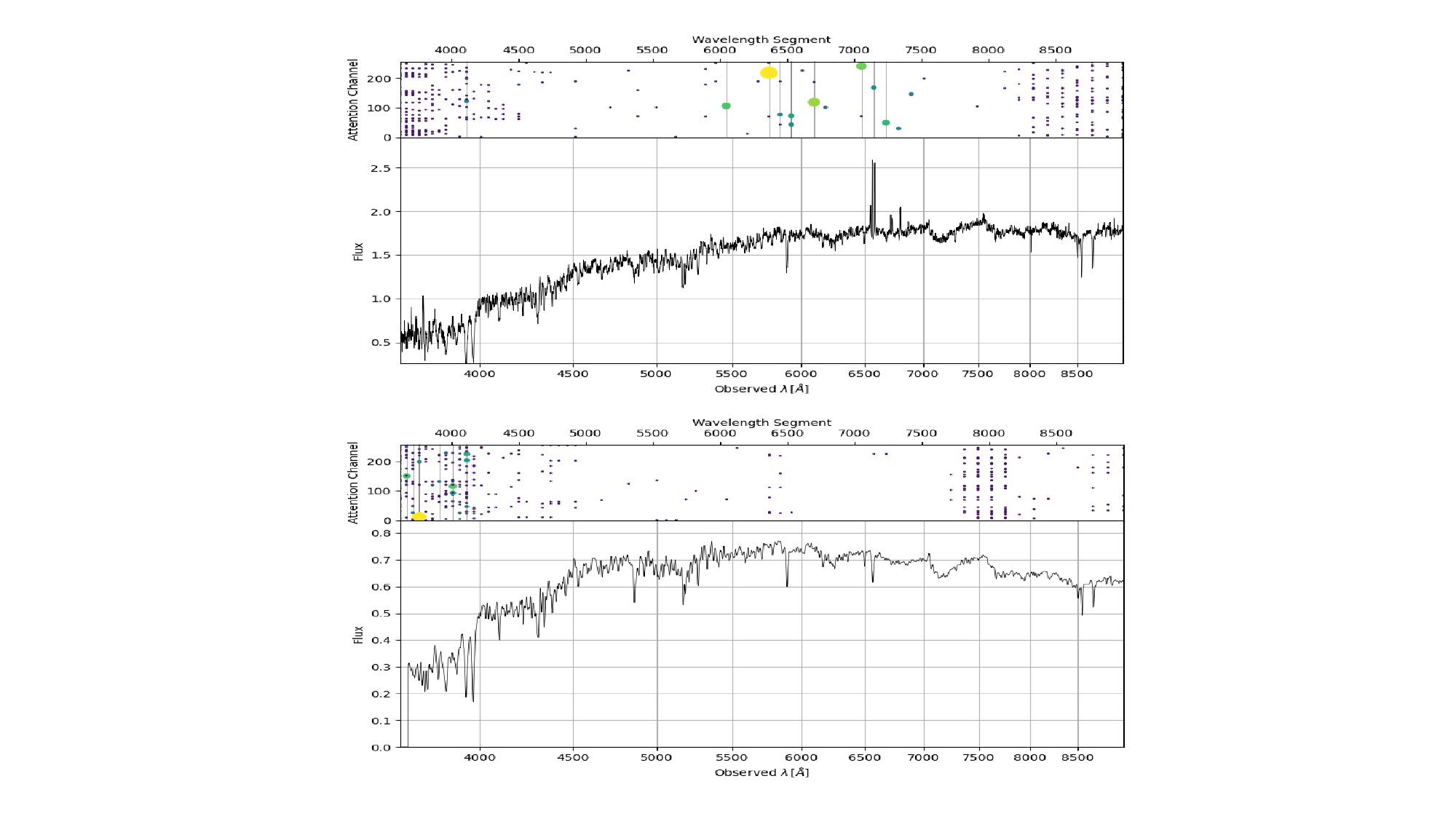}
\caption{Sample spectrum for a LINER classified spaxel with (top) and without (bottom) emission lines. In each figure, the bottom panel is the optical spectrum, and the top panel represents the weights of the attention layer for each channel for each spectral segment. When emission lines are present, most of the attention is focused on them, but with the absence of emission lines the weights are concentrated on the stellar absorption features.}
\label{fig:fig3}
\end{figure}

Given that the spectra in this study all contain strong BPT emission lines, {\sc Spender} could potentially learn their diagnostic ratio patterns and reproduce the same classification scheme. To prevent such redundancy, we train {\sc Spender} on two sets of spectra: one that includes the emission lines and another consisting of only the best-fit continuum model, as available in the Data Analysis Pipeline of MaNGA database. Comparing the results from these two runs provides insights into key features outside of the emission lines and their reliability in classification. Figure~\ref{fig:fig3} illustrates a sample from each of the two sets for a LINER spectrum (as classified by the BPT diagrams), along with the attention weights for each spectrum.

Our sample size for all the classes is equal to the number of data points, with SNR $\geq 5$, within the class with the lowest statistics in the MaNGA database; for our case it happens to be 19,013 LINERs. For training the encoder, we first set aside 15\% of the data for testing. The remaining data is split into an 85\% training set and a 15\% validation set. We train the encoder for 75 epochs using the Adam optimizer \citep{Adam}, the 1Cycle learning rate schedule \citep{Smith2017} with a maximum learning rate of $2\times10^{-3}$, and a batch size of 56. Training on an NVIDIA V100 GPU takes approximately 10 minutes. The final trained encoder exhibits comparable training and validation losses, indicating that the network is not over-fitting and that the training procedure is stable.

Each spectrum is labeled according to its BPT classification, where 0 represents SF regions, 1 denotes Sy, and 2 corresponds to LINERs. We include only spectra that have consistent classifications in the BPT-[NII], BPT-[SII], and BPT-[OI] diagrams; spectra with inconsistent classifications are labeled as "ambiguous" and disregarded. The MaNGA spectra are predominantly composed of SF spaxels, comprising approximately 75\% of the dataset, while only about 2\% are consistently classified as either LINER or Sy. To prevent over-representation of SF spectra during training, we select an equal-sized, randomly chosen sample from each class. Table 1 provides a high-level statistical summary of the data and the detection rates of {\sc Spender} on the test sample.

Within the class of linear encoding methods, the optimal information preserving transformation is given by principal component
analysis (PCA) \citep{Fukunaga+1970}. To benchmark our classification process and asses the performance of {\sc Spender}, we perform our classification scheme using a simple PCA approach. We utilize the Python implementation of PCA from the sklearn library and perform the analysis on the same two sets of spectra (continuum and continuum+lines). We specify the variance (or information) to be retained by the principal components to be 95\% and have the code choose the optimum number of components. We use the same split between the training and test subsets as above for which the optimum number of components was determined to be 2. We fit the algorithm to the training set and then transform both the training and test sets into their principal components. We finally use a Logistic Regression model to map the transformed training data to their designated class, and then make predictions to the test set. The detection rates for both the PCA and {\sc Spender} approaches are reported in Table~\ref{tab:1}. As can be seen, the full spectrum is easily classified by both techniques, and the continuum alone evidently contains information necessary for the classification, but {\sc Spender} performs better especially for the LINER and Sy classes.

\begin{table}
\centering   
\caption{A description for the composition of the dataset, the sample sizes, and the rates at which {\sc Spender} and PCA to successfully classify the test sample.}
\label{tab:1}
     \begin{tabular}[c]{ccc}\hline
      Parameter &  & Value \\ \hline
      S/N threshold &    & $\>5$ \\
      \# spaxels &    & $1.2\times10^6$ \\
      \% SF &    & 75 \\
      \% Sy &    & 2 \\
      \% LINER &    & 2 \\
      \% ambiguous &    & 20 \\ \hline
      Sample size&    & 19,013 \\
      \% Training &    & 85 \\
      \% Test &    & 15 \\ \hline
      &PCA\\ \hline
      Detection & Full & Continuum \\ 
      rate & spectrum & only \\ \hline
      SF &  99\%  & 94\% \\
      Sy &  96\%  & 73\% \\
      LINER &  97\%  & 81\% \\ \hline
      &{\sc Spender} \\ \hline
      Detection & Full & Continuum \\ 
      rate & spectrum & only \\ \hline
      SF &  99\%  & 97\% \\
      Sy &  95\%  & 89\% \\
      LINER &  96\%  & 90\% \\ \hline

    \end{tabular}
\end{table}

\section{Results}\label{section:results}
\subsection{Continuum and Stellar Population}
In the analysis of the detection rates presented in Table \ref{tab:1}, we observe that utilizing the full spectrum for classifying the test sample is significantly more reliable than using the continuum model alone. This outcome is anticipated, as the emission lines incorporated in the full spectrum are consistently employed in BPT diagrams to effectively differentiate between various classes. The encoder demonstrates a strong ability to recognize these emission line patterns, achieving near-perfect detection rates. Additionally, the attention weights, as shown in Fig.~\ref{fig:fig3}, support this conclusion, revealing that the most attention is directed toward the spectral regions containing strong emission lines. Interestingly, there is also a notable cluster of attention weights around the 4000\AA\ region, which is known to contain absorption features linked to the underlying stellar population \citep{Graves+2007}. This suggests that key information about the dominant ionizing source is encoded not only in the emission lines related to the ionized interstellar medium (ISM) but also in the stellar absorption features, which can contribute ionizing photons.

To further investigate, we conducted the same experiment on the spectra with the emission lines subtracted. In this run, nearly all attention weights were concentrated in the 4000\AA\ region (Fig.~\ref{fig:fig3}). Notably, the detection rate for SF spaxels remained almost the same as in the full-spectrum analysis, but there was a slight decrease in the detection rates for Sy and LINER spaxels. This result aligns with expectations, as the ionizing source in SF spaxels—primarily a young stellar population—imprints its signature on both the absorption features and the emission lines excited by these stars. In contrast, Sy and LINER spaxels, where the ionizing photons are presumed to originate from non-stellar sources, should be readily identifiable through their emission lines, with no connection to their stellar absorption features. However, the detection rates for Sy and LINER spaxels were quiet high, indicating that there is a strong connection between the non-stellar ionizing source that is driving the Sy and LINER emission, and the associated stellar population. Or it could be that, for at least a subset of these classes, the stellar population may supply the required ionizing photons, as suggested by \citet{Graves+2007}. To avoid {\sc Spender} learning the residuals after subtracting the lines from the continuum only run, we experiment with replacing these residuals with a flat continuum, and the resulting detection rates are similar to simply subtracting the lines. These results were further validated by repeating the experiment multiple times with randomly selected samples from each class, reinforcing the robustness of the approach. The subsequent sections of this paper will further explore these subgroups of LINER spaxels.

\subsection{Different ionizing sources for LINERs}

\begin{figure}[htp]
\centering
\hspace*{-0.6in}
\includegraphics[trim={2.2in 0.2in 2.5in 0.2in},clip,width=12cm,height=10cm,angle=0]{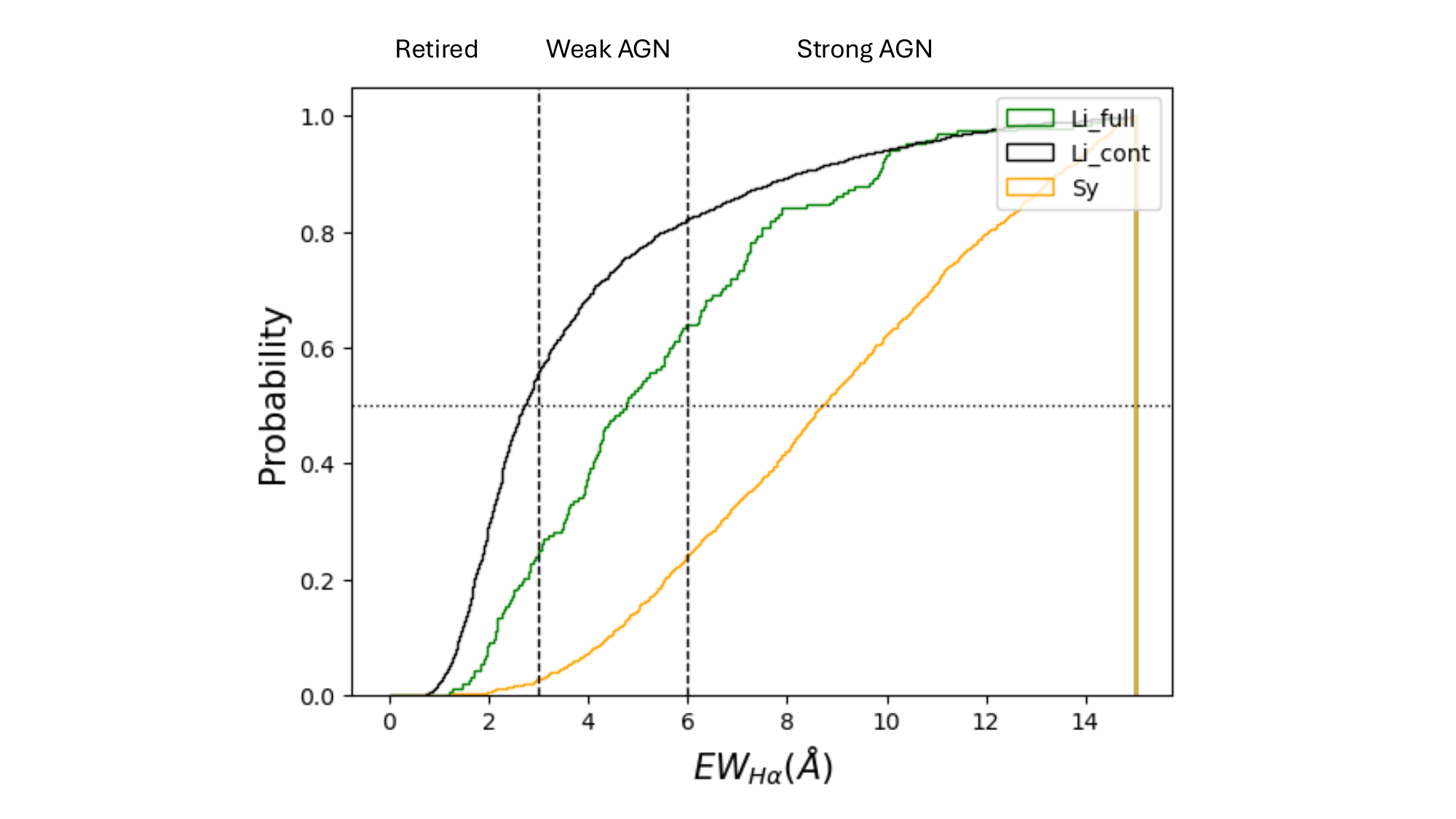}
\caption{A histogram of EW($H{\alpha}$) for $Li_{cont}$, $Li_{full}$, and Sy datasets. The median is depicted with a dotted horizontal line while the limits for the different classes of AGN are indicated by dashed vertical lines. The histogram is normalized so that the area under each of the curves is equal due to unequal sample size. The majority of the $Li_{cont}$ set is determined to be "fake" AGN while the majority of Sy spaxels are strong AGN according to the $H\alpha$ EW.}
\label{fig:fig4}
\end{figure}

In our analysis, we define the subset of LINER spaxels that can be identified solely through the continuum as "$Li_{cont}$", and LINER spaxels whose identification requires both continuum and emission as "$Li_{full}$". This separation allows us to investigate any differences in the ionizing photon sources for each subset. A key insight is provided in Fig.~\ref{fig:fig4}, where we present the Cumulative Distribution Function (CDF) for $H\alpha$ EW for both groups. The $Li_{cont}$ group exhibits lower $H\alpha$ EW values compared to $Li_{full}$, with median values of 2.8 and 5.1, respectively. For reference, we also plot the CDF for the entire Sy population, which displays significantly higher $H\alpha$ EW values with a median well above 6.

This distinction aligns with the classification system proposed by \citet{Fernandes+2011}, where $H\alpha$ EW values of $ \leq 3$ suggest "retired" galaxies or so-called "Fake AGNs", values between 3 and 6 indicate weak AGNs, and values $\geq 6$ are characteristic of strong AGNs. The comparison between $Li_{cont}$ and $Li_{full}$ implies that $Li_{full}$ likely harbors weak AGN activity, whereas $Li_{cont}$ may be dominated by ionizing sources associated with less active or "retired" systems. These findings support the hypothesis that the $Li_{full}$ group is primarily photoionized by weak AGNs, while the $Li_{cont}$ group is more influenced by stellar photoionization processes.

\subsection{Stellar Population Characteristics}

\begin{figure}[htp]
\centering
\hspace*{-1.0in}
\includegraphics[width=22cm,height=12cm,angle=0]{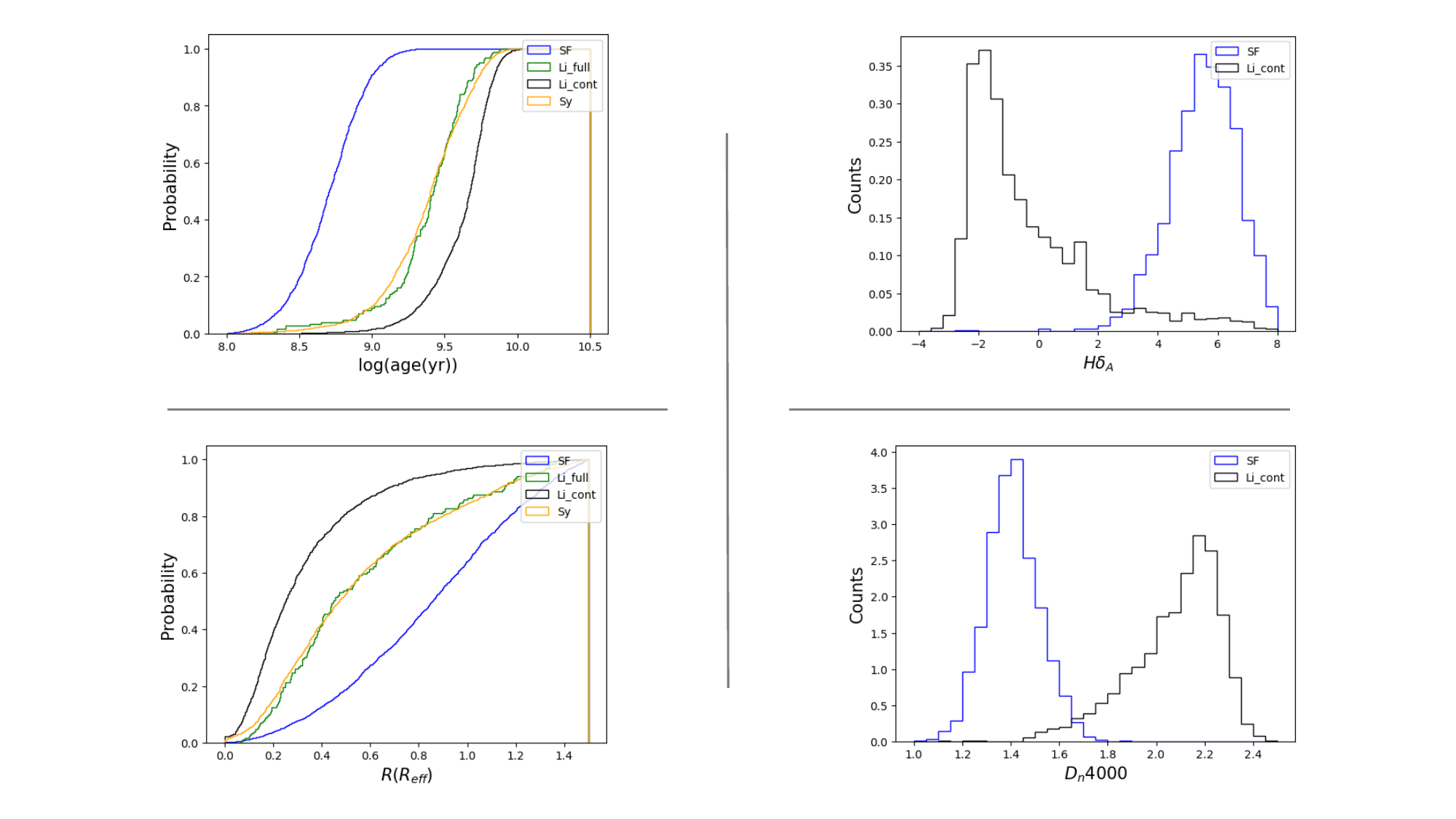}
\caption{Left: CDF for the age of the stellar population (in log(yr)) and the radial distance of spaxels in units of $R_{eff}$ for the same datasets as in Fig.~\ref{fig:fig4} in addition to the SF data. Right: Histograms of Lick indices related to the stellar age. Most of the $Li_{cont}$ spaxels contain an old stellar population residing in the bulge while the Sy and $Li_{full}$ sets have similar age and radial distributions.}
\label{fig:fig5}
\end{figure}

We further investigate the underlying stellar populations for the $Li_{cont}$ subset by analyzing the spectral content of each spaxel using data from the MaNGA Pipe3D Value Added Catalogue (Pipe3D VAC), which provides stellar population parameters, including age and metallicity, derived from full spectral fitting. In Fig.~\ref{fig:fig5}, we present the luminosity-weighted stellar ages for both LINER groups as well as the entire samples of SF and Sy spaxels. For SF spaxels, we observe stellar ages in the range of hundreds of millions of years, consistent with the presence of a young, massive stellar population, which is responsible for supplying ionizing photons in these regions. Conversely, the LINER and Sy spaxels exhibit much older stellar populations, with ages $\geq$1 Gyr, characteristic of an evolved, low and intermediate-mass stellar population, as massive stars do not survive long enough to contribute significantly. Notably, the $Li_{cont}$ group shows the oldest stellar population, with the majority of spaxels in the 3–7 Gyr range. This observation aligns with the models of \citet{Byler+2019}, which show that for a 10 Gyr stellar population, p-AGB stars account for nearly all of the hydrogen and helium ionizing flux.

Additionally, we assess the absorption line strengths in each spectrum using the Lick indices, including the $D_n4000$ and $H\delta_A$ indices, as defined by \citet{Worthey+1997}. The $D_n4000$ index, also known as the ``4000\AA~break'' \citep{Balogh+1999}, arises from a combination of the Balmer break and metal absorption lines in a narrow wavelength range, serving as a sensitive indicator of the ageing of a stellar population. For populations younger than 1 Gyr, $D_n4000$ is relatively insensitive to metallicity, but it becomes more dependent on both age and metallicity for older populations \citep{Graves+2007}. The break becomes more pronounced as cooler stars dominate, while in hotter stars, the metals are highly ionized, reducing the break's significance. On the other hand, the EW of $H\delta_A$ is most sensitive to stars of intermediate spectral types (A to early F), reaching a peak 300–400 Myr after a starburst and then declining with time \citep{Kauffmann+2003}. As shown in Fig.~\ref{fig:fig5}, $Li_{cont}$ spaxels are associated with high $D_n4000$ values and low $H\delta_A$ EWs, ruling out the presence of young stars (younger than 1 Gyr) in these regions, unlike SF spaxels.

Finally, we explore the radial distribution of spaxels, measured in units of effective radius ($R_{eff}$) from the galaxy center, as shown in Fig.\ref{fig:fig5}. The $Li_{cont}$ regions are predominantly central, coinciding with the location of the galactic bulge where the stellar concentration is higher. This is in contrast to the SF spaxels, which are more extended and span a wider range of radii. It is also worth noting that the $Li_{full}$ and Sy groups display nearly identical distributions in both stellar ages and radial distances, suggesting a common origin, possibly linked to AGN activity, as indicated by their $H\alpha$ EW values in Fig.\ref{fig:fig4}.

\subsection{Stellar and Gas Metallicities}
\begin{figure}[htp]
\centering
\hspace*{-0.6in}
\includegraphics[trim={0.15in 1.3in 0.1in 1.3in},clip,width=20cm,height=10cm,angle=0]{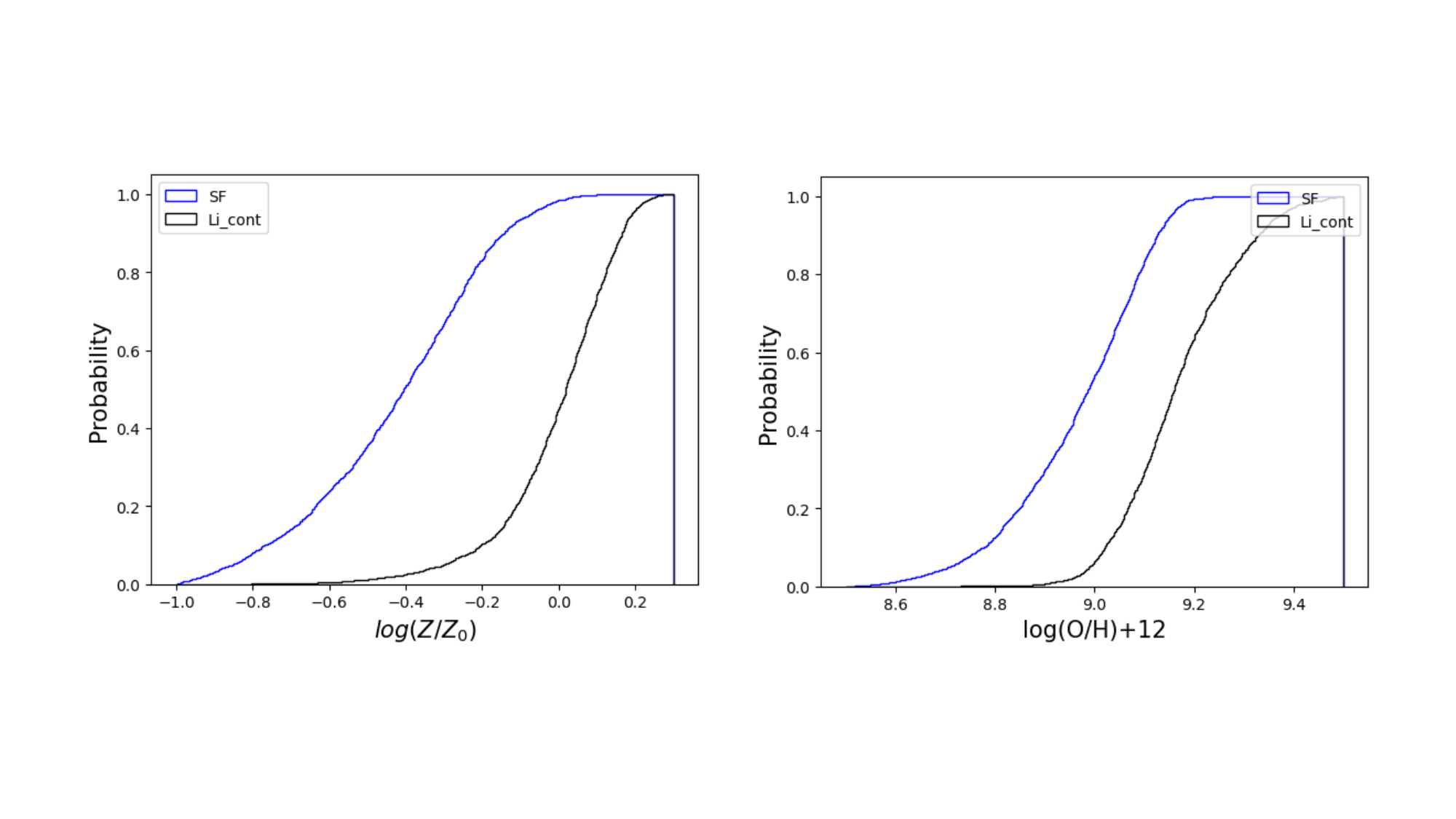}
\caption{Left: A CDF of the oxygen abundance as a proxy for the gas metallicity as inferred using the strong line method for the same datasets as in Fig.~\ref{fig:fig5}. Right: A CDF of the metallicity of the underlying stellar population (in units of ${\rm Z}_\odot$). The stellar and gas components of LINERs have higher metallicity than SF regions indicating internal or external metal enrichment.}
\label{fig:fig6}
\end{figure}

The stellar population's luminosity-weighted metallicity (in units of ${\rm Z}_{\odot}$) is another valuable parameter extracted from the Pipe3D VAC \citep{Lacerda2022NewA...9701895L, Sanchez2022ApJS..262...36S}. We present its CDF in Fig.~\ref{fig:fig6}, comparing the different groups of spaxels. It is noteworthy that the $Li_{cont}$ regions exhibit higher stellar and gas metallicities than the SF regions in our sample. This result is intriguing, as LINERs are typically thought to reside in "retired" elliptical galaxies, as discussed by \citet{Stasinska+2008,Graves+2007,Yan+2006,Belfiore+2016}. Retired elliptical galaxies are sometime assumed to be metal-poor due to the lack of ongoing star formation that would otherwise replenish their heavy elements, although we know that they can also contain ionized gas \citep{Singh+2013}. However, our sample of LINER spaxels reveals a metal-rich stellar population, suggesting that p-AGB stars could be abundant in these regions, as these evolved low-mass stars are significant contributors to heavy metal nucleosynthesis \citep{Cristallo+2015}.

In addition, estimating the gas metallicity in each dataset could provide further insights into the relationship between the emitting gas and the underlying stellar population. We utilize the strong-line method to estimate gas abundances, as outlined by \citet{Kewley+2002}. While this method was developed for HII regions, not p-AGB star environments, the wide range of metallicities and ionization parameters used in the study covers the expected values for p-AGB environments, allowing us to approximate gas abundances, if not precisely, at least relatively. Fig.~\ref{fig:fig6} supports our hypothesis that the gas in $Li_{cont}$ regions is also metal-rich compared to that in SF regions, supporting the idea that the emitting gas is partially enriched by metal-rich ejecta from p-AGB stars.

Moreover, \citet{Byler+2019} conducted one of the most sophisticated recent models of p-AGB stars , and their environments, to explore their relationship to LINER emission. Their models show that p-AGB stars produce distinct emission line ratios compared to young stellar populations, particularly in several emission line diagnostic diagrams. The p-AGB model produces line ratios that fall in the LINER region of the [S\,\textsc{ii}]/$H{\alpha}$ and [O\,\textsc{i}]/$H{\alpha}$ diagrams. However, in the [N\,\textsc{ii}]/$H{\alpha}$ BPT diagram, only models with enhanced $\alpha$ elements, carbon, and nitrogen gas-phase abundances yield LINER ratios, driven by the increased nitrogen content.

\begin{figure}[htp]
\centering
\hspace*{-0.7in}
\includegraphics[trim={0.0in 1.7in 0in 1.7in},clip,width=20cm,height=8cm,angle=0]{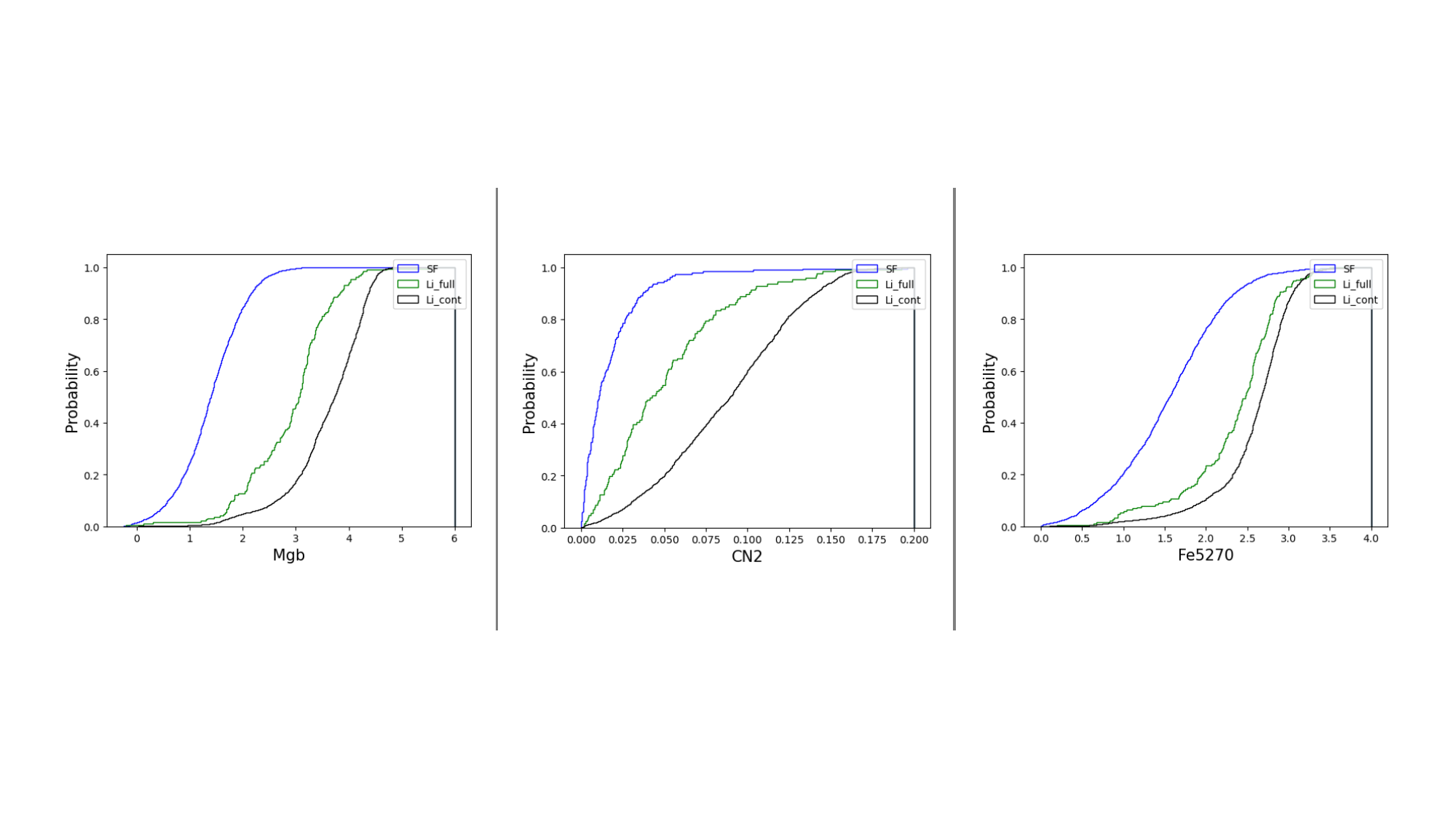}
\caption{A CDF of the Lick indices related to Mg, Fe, C, and N for the underlying stellar population for the same datasets as in Fig.~\ref{fig:fig5}. This suggests metal enrichment in the stellar population of $Li_{cont}$ in C, N and $\alpha$ elements as suggested by \citet{Byler+2019}.}
\label{fig:fig7}
\end{figure}

In Fig.~\ref{fig:fig7}, we examine the Lick indices associated with these elements to verify those findings observationally. The CDF of the Lick indices reveals a pattern where the $CN_2$ and Mg abundances are elevated in $Li_{cont}$ regions compared to $Li_{full}$, and more distinctly compared to SF regions. Interestingly, the Fe index appears similar between the two LINER groups. This suggests an enhancement in [Mg/Fe], as a proxy for $\alpha$ elements enhancement, within the $Li_{cont}$ dataset. Additionally, there is an enhancement in carbon and nitrogen, as suggested by \citet{Byler+2019} to satisfy the conditions for exciting LINER emission according to all the BPT diagrams. These results support the hypothesis where the gas in LINER regions is either composed of AGB star ejecta or has been significantly enriched by low- and intermediate-mass stars.

\subsection{Gas and source coupling}

In this section, we further investigate the kinematic relationship between the gas, its ionizing source, and the stellar component for the $Li_{cont}$ subset. Previous studies \citep{Binette+1994, Belfiore+2016, Yan+2018} suggest that p-AGB stars emit enough ionizing photons to power LINER emission, but for this to occur, the photons must be almost entirely absorbed by the surrounding gas. This implies that the spatial distribution of the ionized gas relative to the ionizing stars should not be as random as seen in SF regions, and the separation between the ionizing source and the gas clouds should be on the order of a few parsecs.

The ionization parameter $(U)$ provides valuable information about the spatial relationship between the gas and its ionizing source. For a radiation-emitting cloud at a distance $r$ from the source, $U(r)$ is defined as the dimensionless ratio of the ionizing photon density to the electron density ($n_e$). Essentially, it represents the density of hydrogen-ionizing photons relative to the total hydrogen gas density \citep{Osterbrock+2006}. In photoionization equilibrium, the ionization parameter is proportional to the ratio between two ionization states of the same element. Within the MaNGA wavelength range, the [O\,\textsc{iii}] $\lambda$5007 to [O\,\textsc{ii}] $\lambda\lambda$3726,28 ratio serves as a reliable proxy for the ionization parameter \citep{Belfiore+2016}. The hardness of the ionizing radiation field ($\alpha$) which describes the power-law slope of the ionizing spectrum, also influences the [O\,\textsc{iii}] $\lambda$5007 to [O\,\textsc{ii}] $\lambda\lambda$3726,28 ratio by modifying the ionization structure of the nebula. Both the [O\,\textsc{iii}]/[O\,\textsc{ii}] and [O\,\textsc{iii}]/$H\beta$ ratios are sensitive to the ionization parameter and the hardness of the ionizing field, though the [O\,\textsc{iii}]/[O\,\textsc{ii}] ratio is more closely tied to $U(r)$ in H II regions and in spectra from harder AGN or old stellar populations \citep{Yan+2012}. To correct for the significant wavelength difference between [O\,\textsc{iii}] $\lambda$5007 and [O\,\textsc{ii}] $\lambda\lambda$3726,28, an extinction correction, using a \citet{Cardelli+1989} extinction law, is applied using the Balmer decrement to accurately measure the [O\,\textsc{iii}]/[O\,\textsc{ii}] ratio. 

\begin{figure}[htp]
\centering
\hspace*{-1.2in}
\includegraphics[trim={0.0in 0in 0in 0in},clip,width=23cm,height=13cm,angle=0]{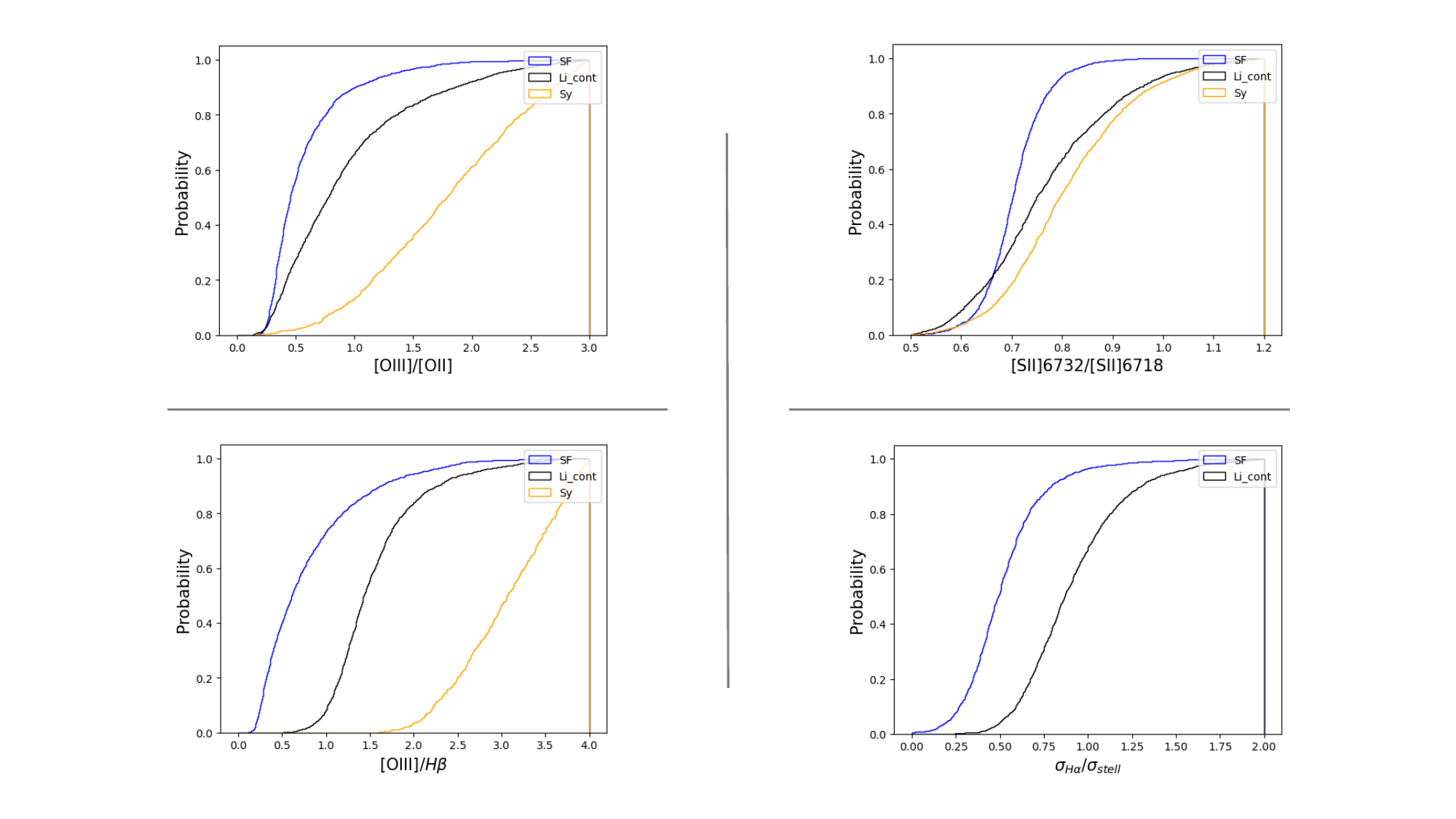}
\caption{CDFs for the [O\,\textsc{iii}]/[O\,\textsc{ii}] lines, [S\,\textsc{ii}] doublet, and [O\,\textsc{iii}]/$H\beta$ for the same datasets as in Fig.~\ref{fig:fig5}. We also show the ratio of ($\sigma_{H\alpha}$) to ($\sigma_\star$). It is clear that LINER regions have harder ionizing spectra, higher ionization parameter, and closer kinematic coupling between the source and the emitting gas than SF regions.}
\label{fig:fig8}
\end{figure}

We see in Fig.~\ref{fig:fig8} that the [O\,\textsc{iii}]/$H\beta$ CDF distribution for the $Li_{cont}$ set is higher than SF, indicating a harder ionization source than young massive stars, yet lower than Sy. Also, the [O\,\textsc{iii}]/[O\,\textsc{ii}] ratio is higher for $Li_{cont}$ regions compared SF, but lower yet Sy indicating a higher ionization parameter $U$ than SF and lower than Sy. Previous work \citep{Binette+1994,Stasinska+2008,Yan+2012} has demonstrated that LINER emission is best reproduced by photoionization by a harder radiation spectrum than that of young stars and a lower ionization parameter than that of Sy AGN. Moreover, The flux ratio of the [S\,\textsc{ii}] $\lambda$6731/[S\,\textsc{ii}] $\lambda$6717, constitutes as direct probe of the gas electron density, with the ratio expected to vary from 0.7 to 2.0,  for $n_e \sim 30$~cm$^{-3}$ to 10$^4$~cm$^{-3}$ while at the lower limit the ratio saturates and can be up to $\pm300$~cm$^{-3}$ uncertain around it \citep{Osterbrock+2006,Delgado+2023}. Indeed, assuming that a single p-AGB star emits enough ionizing photons per second and an electron density of $\sim10^2$~cm$^{-3}$, in agreement with the measured [S II] double ratio in Fig.~\ref{fig:fig8}, an ionization parameter for $Li_{cont}$ higher than SF and is consistent with the scenario where the ionization parameter is high enough to justify the small separation between the gas cloud and the ionizing source in $Li_{cont}$ which is also reported in previous observational studies \citep{Binette+1994,Belfiore+2016,Stasinska+2008,Yan+2018}. 

Finally, we compare the velocity dispersion ratio of $H\alpha$ ($\sigma_{H\alpha}$), which traces the movement of the ISM, to the stellar velocity dispersion ($\sigma_\star$) to assess the kinematic relationship between the gas and stars. The ($\sigma_\star$) is expected to be higher since the stars are approximately collisionless systems while the ISM is prone to friction and continually lose energy and momentum. As seen in Fig.~\ref{fig:fig8}, the ratio for SF spaxels is considerably less than 1, consistent with previous studies \citep{Belfiore+2016}. This is probably because SF regions are born out of cold gas and then ionized and heated by young massive stars. In contrast, $Li_{cont}$ regions exhibit a ratio closer to 1, suggesting that the stellar source shares the same origin as the emitting gas. These results highlight the idea that LINER regions' ionization parameter is higher than typical SF regions, and that the LINER emitting gas is intimately exposed to the photons and ejecta of its stellar population.

\section{Discussion and conclusion}\label{section:discussion}

Our findings that LINER spectra can be reliably identified using their stellar continuum, and that the characteristics of this stellar population are consistent with evolved, low- to intermediate-mass stars, provide new insights into the ionizing mechanisms at play in galaxies. The ability of our machine-learning-based encoder to differentiate LINER spectra from their stellar absorption features alone reinforces the growing body of evidence that evolved stars, particularly p-AGB stars, contribute significantly to the ionization in these regions, challenging the traditional view of AGN as the dominant ionizing source for LINERs. Our analysis reveals that LINER stellar populations are characterized by older ages, higher metallicities, and a central concentration within the galactic bulge. These traits align with previous studies suggesting that p-AGB stars provide the necessary ionizing photons. 

Using the full spectrum for classification, as opposed to the traditional BPT diagrams, demonstrates a significant advantage in capturing the full complexity of galaxy ionization mechanisms. While BPT diagrams rely solely on strong emission lines, our approach incorporates the continuum features, allowing for a more comprehensive analysis of the underlying source. Applying this study to higher redshifts would be a crucial next step in understanding how these ionization mechanisms evolved in the early universe.  Furthermore, investigating these processes at various wavelengths would allow for a more robust confirmation of stellar-driven ionization, particularly in distant galaxies where emission lines are redshifted beyond the optical range.

Finally, the implications of our finding that LINERs, the largest subpopulation of AGN, may not be AGN-driven are profound for our understanding of galaxy evolution. If a significant portion of LINER galaxies are powered by stellar rather than AGN activity, it suggests that nuclear activity in the local universe is far less prevalent than previously thought. This would challenge existing models of galaxy evolution, which often emphasize AGN feedback as a key driver in shaping the growth and behavior of galaxies. The recognition that evolved stellar populations could dominate the ionizing mechanisms in LINER regions highlights the need for a reassessment of the role of AGN in galactic evolution.

\section*{Acknowledgments}

Special thanks to Sultan Hassan and Marc Huertas-Company for their illuminating discussions that greatly improved this work. This material is based upon work supported by Tamkeen under the NYU Abu Dhabi Research Institute grant CASS. Funding for the Sloan Digital Sky Survey IV has been provided by the Alfred P. Sloan Foundation, the U.S. Department of Energy Office of Science, and the Participating Institutions. SDSS-IV acknowledges support and resources from the Center for High-Performance Computing at the University of Utah. The SDSS web site is www.sdss.org.

SDSS-IV is managed by the Astrophysical Research Consortium for the Participating Institutions of the SDSS Collaboration including the Brazilian Participation Group, the Carnegie Institution for Science, Carnegie Mellon University, the Chilean Participation Group, the French Participation Group, Harvard-Smithsonian Center for Astrophysics, Instituto de Astrof\'isica de Canarias, The Johns Hopkins University, Kavli Institute for the Physics and Mathematics of the Universe (IPMU) / University of Tokyo, the Korean Participation Group, Lawrence Berkeley National Laboratory, Leibniz Institut f\"ur Astrophysik Potsdam (AIP), Max-Planck-Institut f\"ur Astronomie (MPIA Heidelberg), Max-Planck-Institut f\"ur Astrophysik (MPA Garching), Max-Planck-Institut f\"ur Extraterrestrische Physik (MPE), National Astronomical Observatories of China, New Mexico State University, New York University, University of Notre Dame, Observat\'ario Nacional / MCTI, The Ohio State University, Pennsylvania State University, Shanghai Astronomical Observatory, United Kingdom Participation Group, Universidad Nacional Aut\'onoma de M\'exico, University of Arizona, University of Colorado Boulder, University of Oxford, University of Portsmouth, University of Utah, University of Virginia, University of Washington, University of Wisconsin, Vanderbilt University, and Yale University.

This work made use of Astropy:\footnote{http://www.astropy.org} a community-developed core Python package and an ecosystem of tools and resources for astronomy \citep{AstropyCollaboration2013A&A...558A..33A, AstropyCollaboration2018AJ....156..123A, AstropyCollaboration2022ApJ...935..167A}.


\end{CJK*}
\end{document}